# Empirically Driven Design of Software Development Processes for Wireless Internet Services

Ulrike Becker-Kornstaedt[1], Daniela Boggio[2], Jürgen Münch[1], Alexis Ocampo[1], and Gino Palladino[3]

[1] Fraunhofer Institute Experimental Software Engineering,
Sauerwiesen 6, 67661 Kaiserslautern, Germany
`{Becker, Muench, Ocampo}@iese.fhg.de`
[2] Motorola GSG-Italy, Via Cardinal Massaia 83, Torino, Italy
ADB005@email.mot.com
[3] Investnet, Via Fava 20, Milan, Italy
`g.palladino@investbv.com`

**Abstract.** The development of software for wireless services on the Internet is a challenging task due to the extreme time-to-market pressure, the newness of the application domain, and the quick evolution of the technical infrastructure. Nevertheless, developing software of a predetermined quality in a predictable fashion can only be achieved with systematic development processes and the use of engineering principles. Thus, systematic development processes for this domain are needed urgently. This article presents a method for the design of an adaptable software development process based on existing practices from related domains, industrial piloting, and expert knowledge. First results of the application of the method for the wireless Internet services domain are described. The benefit for the reader is twofold: the article describes a validated method on how to gain process knowledge for an upcoming field fast and incrementally. Furthermore, first results of the process design for the wireless Internet services domain are given.

## 1. Introduction

Experience indicates that developing software with high quality requirements can only be done successfully if an explicitly defined process is followed. Furthermore, a lack of a development process makes accurate planning very difficult and in many cases impossible. Experience from progressive software development organizations like the NASA Software Engineering Laboratory (SEL) [11], for instance, has shown that one essential precondition for developing software of a predetermined quality in a predictable fashion is the design, establishment, and use of systematic software development processes.

Process deployment can fail especially if an organization does not put enough emphasis into the design and promotion of process models and the infrastructure needed for process deployment. Early results from a multi-case study conducted at Nokia Mobile

Phones clearly show the importance of a stable and well implemented infrastructure for process deployment [24]. A prerequisite for this are explicitly defined process models for the application domain that are tailorable to specific project contexts.

Usually, for new and therefore unknown application domains, no explicitly defined software development processes are available yet. Furthermore, the design and introduction of such processes is very risky, because typically there exists no previous experience on which processes or process fragments are suitable and executable in the environment of the developing organization. An application domain that has to deal especially with such problems is the wireless Internet services domain because its development cycles are very short. In order to produce software of sufficient quality and thus remain competitive in the market, an appropriate and piloted development process is needed very quickly. This is valid in general for a new domain, but it is especially valid for the wireless Internet domain: If a specific process for wireless Internet services is not defined, the risk exists that the process followed in Internet services development will also be inherited for wireless services. As the wireless Internet gets popular, the Internet service providers will try to provide the same services over the wireless Internet as well, and they may easily try to follow the same development process they use for Internet services. This is very risky because the wireless world is different from the fixed world and additional issues must be considered during the implementation of services in order to get a final product with a certain level of quality, which can be competitive on the market.

This article describes a method for the empirical design of development processes for new domains. The overall method can be applied to unknown new domains in general, but as the focus of this work is the wireless Internet domain, special emphasis is placed on the particularities of this domain. First results of the application of the method in this specific domain are discussed. The goal of the method is to rapidly come up with a process that considers existing experience. The process is subsequently evaluated in pilot projects. As a consequence, drastic risk reductions in developing applications for the new domain are expected. The method was applied in the wireless Internet services domain in the context of the WISE project, which involves several European industrial and research organizations.

The two key ingredients for the method are the set-up of selected pilot projects and the creation of descriptive process models from the pilot projects. The pilot projects ought to rely as much as possible on practices already in place in the development organization. For instance, new domains may require new practices or adaptations of existing practices. Variations and commonalities of processes need to be identified. Commonalities may indicate typical process steps and can lead to abstractions of the process in the model; variations are indicators for possible factors impacting the process and may lead to specializations of the process. Finally, the process models are integrated to form a comprehensive process model.

The article is structured as follows: Section 2 gives background information on the problem of process modeling for new domains, introduces the wireless Internet services domain and sketches the WISE project in which this work was done. Section 3 describes the method for designing processes for new domains. First results of the application of this method for the wireless Internet services domain are described in Section 4. A process sketch and an overview of processes and practices from related fields are given. Section 5 briefly surveys existing approaches for designing software development processes. Finally, Section 6 summarizes the article and discusses experiences and open issues.

## 2. Background

An explicit process model is a key requirement for high productivity and software quality. Since software development projects are unique regarding their combination of specific goals and characteristics, providing 'ideal' and at the same time universal development processes is no solution for real life [9]. Instead, effective and efficient software development processes custom-tailored to the particularities of the application domain and project constraints are required. The design of processes for unknown domains implicates several difficulties: 1) Whereas for conventional software development, several standards exist, for new domains no such standards are available that could be used as reference. 2) New domains lack specific experience on particular techniques, their applicability and constraints. 3) The variations of the applications and, as a consequence possible variations of the development processes are not sufficiently understood. 4) The impact of the variation of the enabling technology on the developed service is not always known and this may affect the development process. There are several ways towards solving this problem: one widely accepted idea in the software engineering community is descriptive modeling of development processes, which leads to the explicit definition of process models, product models, and resource models [26]. Descriptive software process modeling attempts to determine processes as they take place in development. Adapting practices and processes from related domains can be a means for getting initial process models.

For establishing baselines (e.g., an effort baseline), collecting and using measurement data may further enhance the understanding and control of software development processes and products, and relationships between them [4]. This leads to the development of empirical quantitative models, which is not the focus of this article.

An upcoming new application domain is the wireless Internet services domain, which can be characterized as follows: rapid disposability of software for wireless Internet services with reasonable quality and high usability has an outstanding importance for the marketability of such services. Wireless Internet services can be characterized by quickly evolving technology, upcoming new devices, new communication protocols, support for new different media types, varying and limited communication bandwidth, together with the need for new business models that will fit in with the completely new services portfolio. Examples of new wireless Internet services can be expected in the domain of mobile entertainment, telemedicine, travel services, tracking and monitoring services, or mobile trading services. At the moment, there is very little experience on developing software for such services systematically. From the viewpoint of a Process Engineer, the following questions arise: How can we quickly adapt software development processes from other domains for the development of wireless Internet services? How can processes be sped up by perpetuating acceptable quality? Which existing techniques, methods and tools can be used? How should these be selected, adapted, and integrated into the process? What are typical variations of the processes in this domain? What are the impact factors on the effects of the processes? What kind of documentation is required?

If a specific process for wireless Internet services development is not soon identified and advertised, the risk exists that experienced developers and content and service providers will apply the same process they succeeded with in developing services for the fixed Internet. This will lead most of them to fail or to produce services that do not fit in with the wireless world requests and cannot turn out to be competitive on the market.

The described work was conducted in the context of the WISE project (Wireless Internet Software Engineering), which was started in 2001 and will run until 2004. The project aims at delivering methodologies and technologies to develop services on the wireless Internet. The methodology part comprises an overall process to drive the engineering of mobile services, a business model to specify roles and skills of involved parties, and guidelines to handle heterogeneous clients (e.g., handhelds, laptops). The technology part comprises a high level architecture for mobile services, a service management component, a data replication and synchronization component, and software agents to support negotiation functions in components. WISE follows an underlying experimental paradigm: Experimenting methodology and technology in real life applications is seen as the key to understanding, validating and improving methodology and technology. Therefore, several pilot developments have already been performed or are planned in the near future. This article describes the results with respect to the software process for the first iteration of two pilot projects. Industrial partners responsible for the pilot development and the underlying infrastructure are Motorola Global Software Group - Italy (Motorola GSG-Italy), Investnet, Sodalia, and Solid. The industrial partners identified several success factors for wireless Internet services, especially time-to-market, the ability to quickly deliver functionality with simultaneous fulfillment of high quality requirements and high usability requirements in terms of service performance. These quality requirements vary with different services. For example, wireless trading services require particularly high reliability, functional correctness, scalability, and redundancy. On the other hand, wireless entertainment has no strict reliability requirements but it has even stricter requirements in terms of usability and performances of the designed service architecture. Research partners responsible for developing processes, methods and tools are the Fraunhofer Institute for Experimental Software Engineering (IESE) in Germany, Politechnico de Torino (Italy), and VTT Electronics (Finland).

## 3. Method

The overall method, its steps along with the major input and output products is depicted in Figure 1. The method consists of the following steps: In the first step, *set-up pilots*, suitable pilot projects have to be determined and organized. Pilot projects are to be determined by market demands in such a way that the pilots are representative for the new application domain. In the step *perform pilots*, the pilot projects are con-ducted. In the step *elicit and model processes*, the processes as performed in the pilot projects are observed and modeled, resulting in a set of descriptive process models. A first version of the process models can be obtained based on similar past projects. The corresponding information is obtained through interviews with involved persons and other information sources, such as project plans or process artifacts.

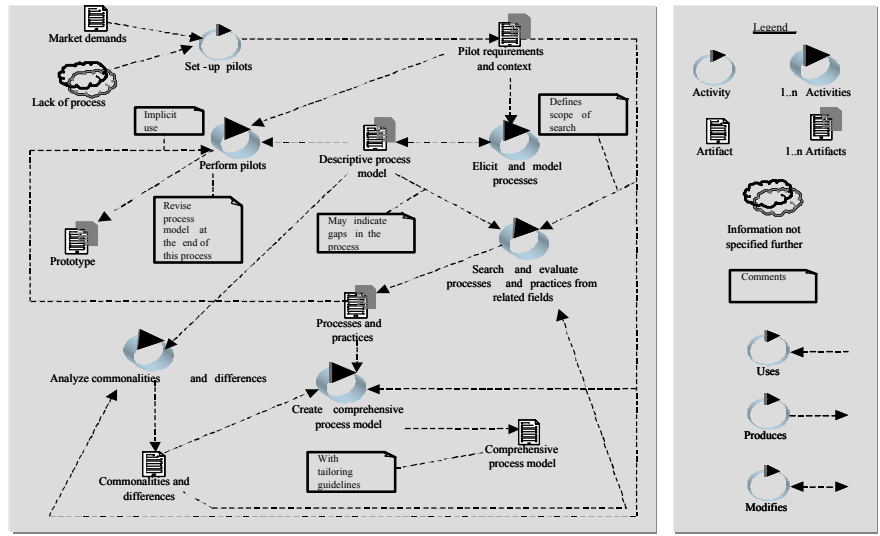

**Fig. 1.** Top level activities of the method

In parallel to these three steps, a step *search and evaluate processes and practices from related fields* is performed: The Process Engineer looks for processes and practices from related areas. This information will be used to fill the process model where it is incomplete and to introduce new practices into the process where old practices were seen as inefficient or are no longer adequate. In the step *analyze commonalities and differences*, commonalities and differences between the different process models have to be analyzed in order to identify process variants and justifications for them. This must recognize differences in the application domain as well as goals and contexts of the pilot developments. In the final step, create *comprehensive process model*, the descriptive models for the pilots, practices and processes from related fields are integrated into a comprehensive process model. Accompanying these steps, continuous improvement of the process during the pilot development with continuous flow of feedback will help to tailor the process during development and identify necessary changes early. The different steps will be detailed in the following sections.

This approach has several benefits: first, performing pilot projects and modeling their processes reveals the strengths and weaknesses of the processes early on. This can be seen as process prototyping. For an organization that introduces a process designed in such a descriptive manner, this reduces potential risks related to the introduction of a newly designed process. Second, introducing a new process based on existing practices typically requires a smaller shift in work procedures and is therefore more likely to be accepted by the Process Performers. Third, this concept allows for an incremental approach, which is more manageable than introducing a process in one shot. An additional benefit of this approach is that Process Performers of the domain are directly involved and can contribute to the development of the new process. Therefore the process is more likely to be accepted and adapted. A bottom-up approach allows to quickly get an accurate model and to avoid problems with theoretical models that do not fit and that are not adequately tailored.

### 3.1 Set up and Perform Pilots

Developing applications for new domains is usually driven by market demands and characterized by a lack of defined processes and experience on how to do it. In order to better understand the processes as well as the application, several pilots should be set up (see Figure 2). Looking at future market requests and what is new and interesting to be investigated by a pilot is a main impact factor on the specification pilot contexts and goals. Additionally, the new domain has to be characterized in order to search for related projects and experience in the developing organization. This characterization is documented within the target context. Afterwards, similar projects are searched and assessed with respect to the reuse potential of practices (i.e., techniques, methods, tools) and processes for the new domain. The result is a set of selected projects. Based on the identified market demands and the experience from the selected projects, the requirements for the pilots and their specific contexts are defined. Besides, the pilots have to be planned and organized.

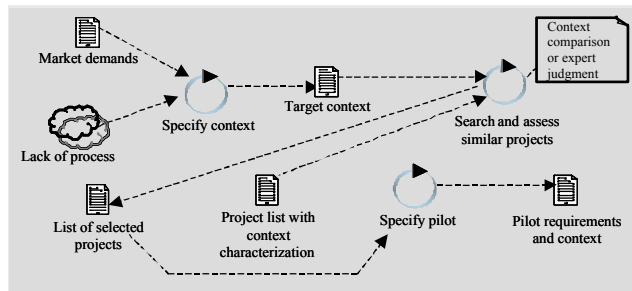

**Fig. 2.** Refinement of s*et up pilots*

In the step *perform pilots* the pilot projects are conducted. Pilot performance is both, product and process prototyping. It combines the benefits of product prototyping (early validation of requirements, understanding the technology, experiencing the architecture etc.) with the benefits of process prototyping (evaluating procedures and practices, understanding the effects of processes, identifying organizational problems etc.).

### 3.2 Elicit and Model Process

For each pilot project, a descriptive process model is developed. Figure 3 details the elicitation and modeling activity. For the identification of existing processes, we recommend the Prospect [7] approach to descriptive process modeling: The main information sources used are interviews with Process Performers and analysis of documents used or produced in the process. The identification of existing processes consists of two stages, *orientation* and *detailed elicitation*. During the orientation phase a process outline is developed. The process outline provides an overview of the process and facilitates further elicitation activities. For example, process information can be described with the help of the process modeling schema [29] implemented in the Spearmint [6] tool. The outline will help sample interviewees and select information sources in the second stage, *detailed elicitation*. If weaknesses in the current process are already known, they should be eliminated. Thus, during the interviews Process Performers should already be asked which practices in the current process they consider inefficient.

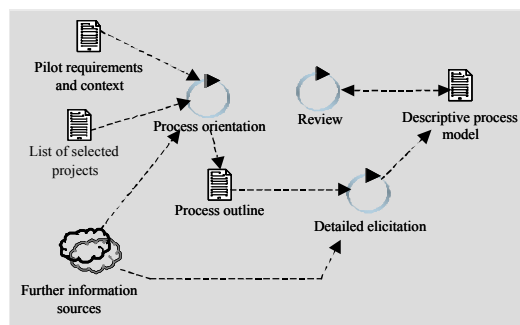

**Fig. 3.** Refinement of e*licit and model process*

Subsequently, the process model is reviewed: People who provided information for the model are asked to review the model to make sure that all information captured was correctly transformed into the model. The result is a description of pilot processes as they actually are being performed in the respective environment. One further benefit of this method is that involving Process Performers early increases process awareness among them. Moreover, involving Process Performers in the definition/tailoring of the process can lead them to more strictly follow a process that they somehow have helped to define, rather that a process that is externally defined and imposed.

### 3.3 Analyze Models and Create Comprehensive Process Model

The different process models are integrated to form a comprehensive model. To develop a comprehensive process model, commonalities and differences among and between the different process models have to be analyzed. Commonalities between the pilot processes may indicate typical process steps and can lead to abstractions of the pilot process models. Variations of the pilot processes may lead to specializations of the comprehensive process model. In this case, different context characteristics (such as developers' experience, system type) of the pilots may indicate the reasons for process variations. If processes differ and no context deviations can be identified, the context is probably not characterized completely and there is at least one influence factor on the process that has not been identified yet. The comprehensive model comprises guidelines or rules on how to adapt the generic parts of the comprehensive model to project-specific goals and characteristics.

As a prerequisite, an appropriate representation of the comprehensive process model and tailoring mechanisms is required. Identification of specific conditions and effective adaptations can be done based on process designer experience, on continuous feedback from process performers during service implementation, and on available literature and historical data on similar project. Several software reuse approaches (e.g., templates, generation, composition, transformation) can be applied to express the comprehensive process model and the tailoring mechanism. One example approach for describing and tailoring comprehensive models is the ProTail approach [5], which is based on a formal process modeling language and a transformational tailoring technique.

### 3.4 Search and Evaluate Processes and Practices from Related Fields

This step comprises the search for information that can be relevant for the design of development processes for the new domain. This step is done in parallel to the steps *set-up pilots*, *perform pilots* and *elicit and model process*. The pilot requirements and contexts identified during the set-up step are input for the definition of the scope of the survey. The descriptive models of the pilot processes might indicate gaps, i.e. process steps that are not clearly understood, situations where the procedure on how to proceed is unclear or experience is required, or gaps in technical knowledge. These gaps might be filled with processes or practices from related fields. During the performance of the pilots, identified practices and processes for gaps are used implicitly. Afterwards, they should be explicitly integrated into the comprehensive model, if their performance was successful.

## 4. Application and First Results

This section describes the application of the method for two pilots. In the WISE project, pilots are a means for designing processes and understanding the technology and methodology to engineer and operate with wireless Internet services in realistic contexts and different application domains. Based on market demands (such as the need to adapt existing services for the Internet towards wireless Internet services or to create new services) and companies' interests, initially two target contexts for the two pilots were defined: the development of a wireless Internet service for mobile online trading (Pilot 1) and the development of a service for mobile entertainment (Pilot 2). An excerpt of the description of the pilot contexts is shown in Table 1.

**Table 1.** Context description for the pilots

| Characteristics | Pilot 1 | Pilot 2 |
|---|---|---|
| Application domain | Service development/mobile online trading | Service development / mobile entertainment |
| Project type | System adaptation | Creation from scratch |
| System type / component type | Application software | Application software |
| Experience of developers | Professional developers | 8 professional developers, 1 student |
| Domain analysis technique | Informal | Informal (provided by domain experts) |
| Requirements technique | List in natural language, intended screen masks, forms, and outputs | Structured text / UML use cases |
| Design technique | UML state diagrams, UML sequence diagrams, UML package diagrams, WISE-specific component diagrams | UML state diagrams, UML sequence diagrams, UML package diagrams, UML class diagrams, WISE-specific component diagrams |
| Implementation technique | WML | Java on both client side (J2ME) and server side (J2EE) |
| Validation technique | Black-box testing | White-box unit testing with JUnit tool, different integration testing techniques, feature testing directly on the target terminal |
| Organizational context | Investnet | Motorola GSG-Italy and Sodalia |

The goal of Pilot1 is to provide a service for the management of a virtual portfolio. For Pilot 1, similar projects could be identified that are concerned with the development of Internet trading services (i.e., development of a market informational and trading simulator site). Pilot 1 is an adaptation of this service to the wireless domain. The requirements for Pilot 1 comprise very high availability, correctness of data, and stringent reliability of customer identification and authorization as well as instantaneous response time in terms of quick data provision.

The goal of Pilot 2 is the development of a multiplayer online game operated from mobile terminals. The requirements for Pilot 2 comprise the ability for user interaction on a shared environment, short response times and the portability on different platforms. For Pilot 2, no similar projects with regard to the application domain were identified because the intention is to develop a completely new service. Nevertheless, existing similar projects concerned with the development of server and client software could be identified.

First results of the descriptive process modeling, the analysis of commonalities and differences, and the search and evaluation of processes and practices from related fields are described in the subsequent sections. The creation of a comprehensive process model can be based on these results. The results are based on the first iteration of the pilot projects. It is planned to have three iterations of each pilot: In the first iteration, the pilots are built with a very sketchy version of methodology and technology. A second iteration uses enhanced methodology and technology as well as an enhanced underlying standard (such as UMTS). The third iteration uses a consolidated version of technology and methodology. In order to gain experience, the pilots are performed with an accompanying goal-oriented measurement program that provides information feedback to all parties and helps in controlling and understanding processes and products and identifying cause-effect relations between them.

### 4.1 Process Sketch

This section sketches the initial process models for each pilot, and lists most striking commonalities and differences found between the pilots. Specifics of the pilot process can be found on the refined level: For example, the development phase of Pilot 1 contains the following list of activities: develop prototype, create preliminary system, release system preliminary, rework. These activities seem to be strongly related with the requirements of the wireless Internet services domain. The development of a prototype and preliminary system helps to achieve and demonstrate part of the application's functionality early in a project. Also, it will be useful to reduce risks with new technical requirements. Furthermore, appropriate network infrastructure has to be established and a friendly customer needs to be involved in the validation of the preliminary system. The product flow of the process model for Pilot 2 (mobile entertainment) is shown in Figure 4. The Figure does not include the description of the control flow, i.e., the performance sequence.

Activities, artifacts, roles, and tools from every pilot are analyzed to obtain valuable information for the comprehensive process model. Some very striking commonalities detected between the two pilot processes are the following: Both pilots have very close involvement by the customers and the providers of technical infrastructure. Market demands need to be carefully examined and understood. In both processes, a design document is produced as input for the implementation phase. Commonalities can be especially recognized in the structure of the architecture, which is described in the design document. For instance, logical architecture components for user authentication, billing, accounting, and user profiling can be found in both architectures of the pilots. This has implications on the processes, e.g., the development of respective interfaces has to be considered by the processes. There is at least one activity for explicitly setting up the test environment in both pilots. In both pilots, there are internal tests as well as external tests with a provisional technical infrastructure. Finally, both pilot processes include an acceptance test in the customer environment.

Some of the differences that were encountered are the following: Pilot 2 is a new development oriented to a wider market spectrum, which implies to cover more platforms, while Pilot 1 is an adaptation of an existing service towards the wireless do-main. The main adaptation tasks in Pilot 1 are downscaling of functionality and interface adaptations. Therefore, the focus of testing in Pilot 1 is more on the interface and less on testing single units. Furthermore, Pilot 1 uses WAP browsers on the client side, whereas Pilot 2 has to develop a dedicated client. Other process differences result from different user interface requirements (developing multimedia interfaces requires other procedures than developing pure textual interfaces) and different non-functional requirements (e.g., mobile online trading requires much higher reliability, which can be supported by implementing redundancies on the server side, for instance). This results, for example, in different validation tests for the final product: Pilot 1 shall be reliable and guarantee highly secure transactions, Pilot 2 shall guarantee high performances and usability features, therefore different test suites shall be designed to verify that the final products fulfill specific requirements.

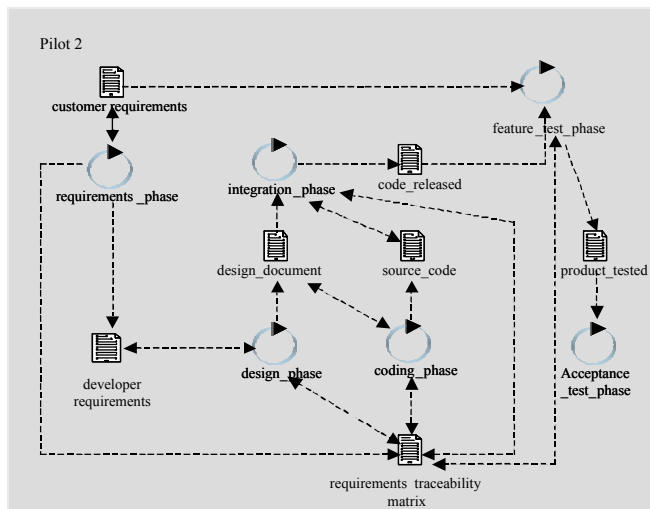

**Fig. 4.** Top-level product flow for Pilot 2

### 4.2 Existing Processes and Practices from Related Fields

This section sketches the search for information that can be relevant for the design of new software development processes for wireless Internet services. There was no literature available that describes software development process models as a result of case studies, surveys or experiments for the wireless Internet services domain as a whole. This emphasizes the importance of the WISE project goal.

The process models in related domains found (Adamopoulos *et al.* [1]), (Karlsson [16]), (Gutowsky [12]), (Zettel *et al.* [30]), (Nilsson *et al.* [23]) are based on an incremental development software process model. Whether the application is for the Internet or the telecommunications domain, it should be delivered as soon as possible, and increments are suggested to maintain control. Therefore, it is probable that for the wireless Internet services domain an incremental development approach is important.

Industrial key players in the telecommunications domain (Gutowski [12]), (Nilsson *et al.* [23]), and (Tapani [19]) have developed process models as part of a software process improvement program. The use of metrics was essential to control, observe and publish the progress on the models. Therefore, a metrics program becomes a priority for any process model to be established.

Two process models (Adamopoulos *et al.* [1]) and (Zettel *et al.* [30]) use as explicit notation UML and Spearmint respectively, to describe activities, inputs, roles and outputs. The need of a simple, generic, understandable notation will be relevant for easily describing the process model for wireless Internet services.

There exist guidelines and practices for web site content development and layout design (Taylor *et al.* [27]). They range from simple ones like a navigation chart, to more sophisticated ones like 'usage-centered design'. The papers found (Nerurkar [22]), (Hammar [13]), (Roe *et al.* [25]), (Constantine *et al.* [10]) point out the 'usability' of the web sites as the key to success in business. The same can be expected for the wireless Internet services domain, and with new hardware constraints for the handheld devices and mobiles small screens. This should give more priority to user interface design research and practices in the future.

It is very clear that even though there is a lot of information published about tools and recent findings on Internet, wireless networks, and mobile computing technologies, there is no published information on established, explicit process models, or guidelines for developing software for the wireless Internet services domain. It is a new field where case studies and academic experiments should be documented in order to establish a knowledge base.

## 5. Related Work

Several papers report their experience on the design of processes. Several of these experience reports deal with redesign or re-writing of official process handbooks in a more formal notation (see for instance [3] or [21]).

Arlow *et al.* [2] describe how a class library management process at British Airways was modeled at a very fine-grained level. In this organization a complete picture of the overall process was missing. Thus, the goal of the modeling task was to develop a comprehensive process model. From the official process documentation, information regarding roles, responsibilities, and library structure could be elicited. From this information, process models using state transition diagrams – a notation Process Performers were familiar with – were developed. The diagrams were discussed in interviews with Process Performers.

Kellner and Humphrey created a process description based on existing practices [15]. They used information gained from interviews with the people directly involved in process execution. This information was supplemented with information gained from people involved in managing the process, and with regulations on the process.

Henry and Blasewitz [14] describe their experience in developing a common process at General Electric Aerospace. As the process involves several organizations, and different groups within these organizations, the main goal here was to obtain a common and consistent overall picture of the process. Each of these groups was tasked to define their process phase. The different models were then validated to ensure consistency. Then a series of meetings with representatives of each group was conducted to review and finalize phase definitions.

In general, it can be said that none of these approaches developed a process for a new application domain, but all of the experience reports listed above focus on the formalization or description of processes that are already in place and have been around for a while. Thus, the steps described in these experience reports can be compared to the step *elicit and model processes*. However, none of these describes how to design a process for a new and unknown domain.

## 6. Summary and Discussion

This article presented a method on how to gain process knowledge for an upcoming application domain and surveyed first results of the process design for the wireless Internet services domain. Several experiences with the approach have been made: The performance of pilots helps to avoid later problems with respect to process and product. The time spent on the pilots is probably much less than the time needed for fixing problems in the process - had the process not been piloted. The process elicitation approach used has been proven effective. Descriptive modeling helped recognize and react on weaknesses in the process very early. Additionally, recommendations for process improvements from the developers could be considered. The review of the process models was difficult, because the developers were located in different geographical locations. An electronic process guide (EPG) helped improve the review. Coupling an EPG with an off-line commentation system would improve the review procedure further. During pilot performance it is necessary to verify that the descriptive process and the process actually performed do not deviate. It is recommended to perform a measurement program in addition to the pilots. This will allow to get initial effort and defect baselines as well as a deeper understanding of the success factors (such as time-to-market) and their influences (such as requirements stability). Guidelines for such a measurement program are, for example, provided in [8].

Summarizing the experience with the wireless Internet services domain, it can be said that the main impact factors on process design are the necessity to understand varying market demands and technology changes, as well as a set of specific nonfunctional requirements for wireless Internet services. Other important characteristics of the domain are that user interface design plays an important role and that it is difficult to set up an appropriate test environment. These specifics have to be considered in the process design.

Future work will be the performance of pilot iterations and the creation of a comprehensive process model. A third pilot project for a wireless content downloading service is starting at this moment at Solid and Sonera. The presented process design approach should be evaluated further in other new domains. Nevertheless, the initial results for the wireless Internet services domain show that the presented process design approach helps to rapidly come up with an approved process model that drastically reduces development risks.

*Acknowledgements*. The work has been funded by the European Commission in the context of the WISE project (No. IST-2000-30028). We would like to thank the WISE consortium, especially the coordinator Maurizio Morisio, for the fruitful cooperation. Additionally, we would like to thank Filippo Forchino from Motorola GSG-Italy for the valuable comments to this article, and Sonnhild Namingha from the Fraunhofer Institute for Experimental Software Engineering (IESE) for reviewing the first version of the article.

# References

[1] Adamopoulos, D.X., Pavlou, G., Papandreou, C.A.: An Integrated and Systematic Approach for the Development of Telematic Services in Heterogeneus Distributed Platforms. Computer Communications, vol. 24, pp. 294-315 (2001)

[2] Arlow, J., Bandinelli, S., Emmerich, W., Lavazza, L.: A Fine-grained Process Modelling Experiment at British Airways. Software Process–Improvement and Practice, vol. 3, No 3., pp. 105-131 (1997)

[3] Aumaitre, J.M., Dowson, M., Harjani, D.R.: Lessons Learned from Formalizing and Implementing a Large Process Model. In: Warboys, Brian., (ed.): Proceedings of the Third European Workshop on Software Process Technology, pp 228-240. Lecture Notes in Computer Science vol. 772. Springer–Verlag, Berlin Heidelberg New York (1994)

[4] Basili, V.R., Rombach, H.D.: The TAME Project: Towards Improvement-Oriented Software Environments. IEEE Transactions on Software Engineering, vol. 14, No. 6, pp. 758-773 (1988)

[5] Becker-Kornstaedt, U., Hamann, D., Münch, J., Verlage, M.: MVP-E: A Process Modeling Environment. IEEE Software Process Newsletter vol. 10, pp. 10-15 (1997)

[6] Becker-Kornstaedt, U., Hamann, D., Kempkens, R., Rösch, P., Verlage, M., Webby, R., Zettel, J.: Support for the Process Engineer: The Spearmint Approach to Software Process Definition and Process Guidance. Proceedings of the Eleventh Conference on Advanced Information Systems Engineering (CAISE '99), pp. 119-133. Lecture Notes in Computer Science, Springer-Verlag. Berlin Heidelberg New York (1999)

[7] Becker-Kornstaedt, U.: Towards Systematic Knowledge Elicitation for Descriptive Software Process Modeling. In: Bomarius, F., Komi-Sirviö, S., (eds.): Proceedings of the Third International Conference on Product-Focused Software Processes Improvement (PROFES). Lecture Notes in Computer Science, vol. 2188, pp. 312-325. Springer-Verlag. Berlin Heidelberg New York (2001)

[8] Briand, L.C., Differding, C., Rombach, H.D.: Practical Guidelines for Measurement-Based Process Improvement. Software Process. Improvement and Practice, vol. 2, No. 4, pp. 253-280 (1996)

[9] Brooks, F.P. Jr.: The Mythical Man-Month. Essays on Software Engineering, Anniversary edition. Addison Wesley. Reading MA (1995)

[10] Constantine, L., Lockwood, L.: Usage-Centered Engineering for Web Applications. IEEE Software, vol. 19, No. 2, pp.42-50 (2002)

[11] McGarry, F., Pajerski, R., Page, G., Waligora, S., Basili, V.R., Zelkowitz, M.V.: An Overview of the Software Engineering Laboratory. Software Engineering Laboratory Series Report, SEL-94-005, Greenbelt MD USA (1994)

[12] Gutowski, N.: An Integrated Software Audit Process Model to Drive Continuous Improvement. Proceedings of the 8th international conference on software quality, pp. 403-415. Portland USA (1998)

[13] Hammar, M.: Designing User-Centered Web Applications in Web Time. IEEE Software, vol. 18, No. 1, pp. 62-69 (2001)

[14] Henry, J., Blasewitz, B.: Process Definition: Theory and Reality. IEEE Software, vol 9, pp. 103-105 (1992)

[15] Kellner, M., Hansen, G.: Software Process Modeling: A Case Study. In: Proceedings of the 22nd Annual Hawaii International Conference on System Sciences, vol. II, pp. 175-188 (1989)

[16] Karlsson, E.: A Construction Planning Process. Q-Labs, LD/QLS 96:0381, Lund Sweden (1999)

[17] Karlsson, E., Vivaldi, N., Urfjell, T.: Guidelines for Step-Wise Design. Q-Labs, LD/QLS, 95:0520, Lund Sweden (1999)

[18] Karlsson, E., Taxen, L.: Incremental Development for AXE 10. ACM SIGSOFT Software Engineering Notes, vol. 22, No. 6 (1997)

[19] Kilpi, T.: Implementing a software metrics program at Nokia. IEEE Software, vol. 18, No. 6, pp. 72-77 (2001)

[20] Kovari, P., Acker, B., Marino, A., Ryan, J., Tang, K., Weiss, C.: Mobile Applications with Websphere Everyplace Access Design and Development. IBM SG24-6259-00 (2001)

[21] Krasner, H., Terrel, J., Linehan, A., Arnold, P., William, H.: Lessons Learned from a Software Process Modeling System. Communications of the ACM, vol.35, No. 9, pp. 91-100 (1992)

[22] Nerurkar, U.: Web User Interface Design: Forgotten Lessons. IEEE Software, vol. 18, No. 6, pp. 69-71 (2001)

[23] Nilsson, A., Anselmsson, M., Olsson, K., Johansson, Erik.: Impacts of Measurement on an SPI Program. Q-Labs (http://www.q-labs.com/files/Papers/SPI99_Imp_of_Meas_on_SPI.pdf)

[24] Raffo, D., Kaltio, T., Partridge, D., Phalp, K., Ramil, J.F.: Empirical Studies Applied to Software Process Models. In: International Journal on Empirical Software Engineering, vol. 4, No. 4 (1999)

[25] Roe, C., Gonik, S.: Server-Side Design Principles for Scalable Internet Systems. IEEE Software, vol.19, No. 2, pp. 34-41 (2002)

[26] Rombach, H.D., Verlage, M: Directions in Software Process Research. Advances in Computers, vol. 41, pp. 1-63 (1995)

[27] Taylor, M.J., McWilliam, J., Forsyth, H., Wade, S.: Methodologies and Website Development: A Survey of Practice. Information and Software Technology, pp. 381-391 (2002)

[28] Upchurch, L., Rugg, G., Kitchenham, B.: Using Card Sorts to Elicit Web Page Quality Attributes. IEEE Software, vol. 18, No. 4, pp. 84-89 (2002)

[29] Webby, R., Becker, U.: Towards a Logical Schema Integrating Software Process Modeling and Software Measurement. In: Harrison, R. (ed.): Proceedings of the Nineteenth International Conference on Software Engineering Workshop. Process Modeling and Empirical Studies of Software Evaluation, pp. 84-88 Boston USA (1997)

[30] Zettel, J., Maurer, M., Münch, J., Wong, L.: LIPE: A Lightweight Process for E-Business Startup Companies based on Extreme Programming. Proceedings of the Third International Conference on Product-Focused Software Processes Improvement (PROFES), pp. 255-270, (2001)